\documentclass[prb,twocolumn,amsmath,amssymb,superscriptaddress]{revtex4}
\usepackage{graphicx}
\usepackage{dcolumn}
\usepackage{bm}

\begin{document}

\title{Infrared phonon activity in pristine graphite}

\author{M. Manzardo
\footnote{Present address:
Leibniz-Institut f{\"u}r Festk{\"o}rperphysik
und Werkstofforschung, Helmolzstra\ss e 20, Dresden, Germany
}}
\affiliation{Dipartimento di Fisica,
``Sapienza'' Universit\`a di Roma, P.le A. Moro 2, 00185 Roma, Italy}

\author{E. Cappelluti}
\affiliation{Instituto de Ciencia de Materiales de Madrid,
CSIC, 28049 Cantoblanco, Madrid, Spain}
\affiliation{ISC-CNR, via dei Taurini 19, 00185 Roma, Italy}

\author{E. van Heumen
\footnote{Present address:
van der Waals - Zeeman Institute, University of Amsterdam,
Sciencepark 904, 1098 XH Amsterdam, the Netherlands
}}
\affiliation{DPMC,
Universit\'{e} de Gen\`{e}ve, 1211 Gen\`{e}ve,
Switzerland}

\author{A.B. Kuzmenko}
\affiliation{DPMC,
Universit\'{e} de Gen\`{e}ve, 1211 Gen\`{e}ve,
Switzerland}

\begin{abstract}
We study experimentally and theoretically the Fano-shaped phonon peak at 1590
cm$^{-1}$ (0.2 eV) in the in-plane optical conductivity of pristine graphite.
We show that the anomalously large spectral weight and the Fano asymmetry of
the peak can be qualitatively accounted for by a charged-phonon theory. A
crucial role in this context is played by the particle-hole asymmetry of the
electronic $\pi$-bands.
\end{abstract}

\date{\today}


\maketitle

The discovery of graphene has triggered a tremendous amount of activity to
study the electronic and structural properties of single-layer and
multi-layer graphitic compounds. The presence, in optical probes, of phonon
anomalies related to the in-plane carbon displacements has in particular been
useful both for the characterization and for the investigation of the
fundamental properties of graphene. While initially Raman spectroscopy was
mainly employed,\cite{pisana,yan,bonini,malard,yan2,gava,malard2} recent
infrared experiments in bilayer and multilayer graphene have significantly
contributed to the field, showing new remarkable phenomena, such as a strong
modulation of the magnitude and of the lineshape of the optical phonon peaks
as function of electrostatic gating,\cite{noi,tang} the number of layers,\cite{li}
and chemical doping.\cite{liu}

The interest to graphene has also given a new impulse to the research on its
3D parent material, graphite. One of the long standing problems is the origin
in graphite of the infrared phonon activity in the optical conductivity at
$\omega \approx 0.2$ eV, which is associated with the in-plane $E_u$ carbon
displacements. Early measurements of the peak intensity, expressed in terms
of the effective infrared charge $Z$, range from $Z=0.18$ to $Z=0.41$
electrons per atoms,\cite{nemanich,dresselhaus} qualitatively confirmed by
more recent measurements.\cite{tongay,humlicek} The possible origin of the
infrared activity of the $E_u$ mode in pristine graphite was theoretically
investigated in Ref. \onlinecite{mahan} by means of a generalized bond-charge
model. The predicted phonon intensity was however significantly smaller than
the experimental results.

In this work we combine experiment and theory in order to gain new insights
into this problem. Our infrared measurements on highly
ordered pyrolytic graphite (HOPG) \cite{kuzmenko2} allowed us to characterize both the intensity
and the Fano asymmetry of the phonon anomaly. We analyze these results within
the context of the charged-phonon theory,\cite{rice} which has recently been
shown to account quantitatively for the phonon infrared intensity in
few-layer graphenes.\cite{li,capp} While in gated graphene the main factors
that determine the infrared intensity are the induced doping and the internal
electric field, here we show that in charge neutral graphite the phonon
infrared activity is essentially driven by the tight-binding terms breaking
the particle-hole symmetry.

\begin{figure}[b]
\includegraphics[width=8.5cm,clip=]{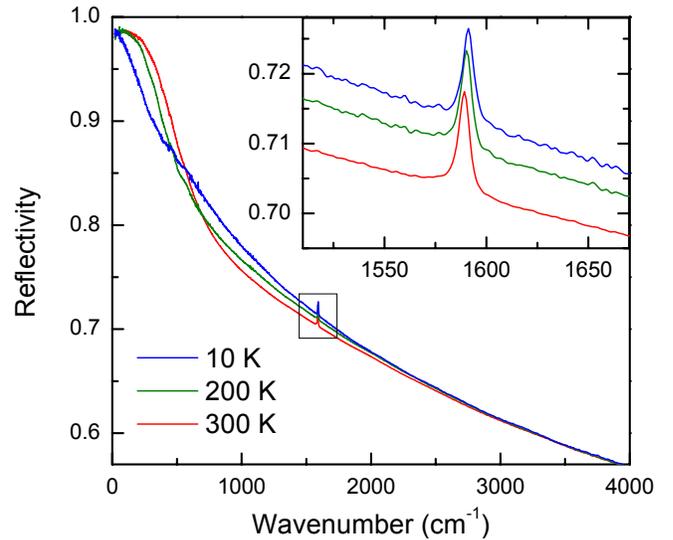}
\caption{Infrared reflectivity spectra of highly ordered epitaxial graphene
at 300 K, 200 K and 10 K from Ref.\onlinecite{kuzmenko2}. The inset shows the region near the phonon peak.}
\label{fig:fig1}
\end{figure}

\begin{figure}[t]
\includegraphics[width=8.5cm,clip=]{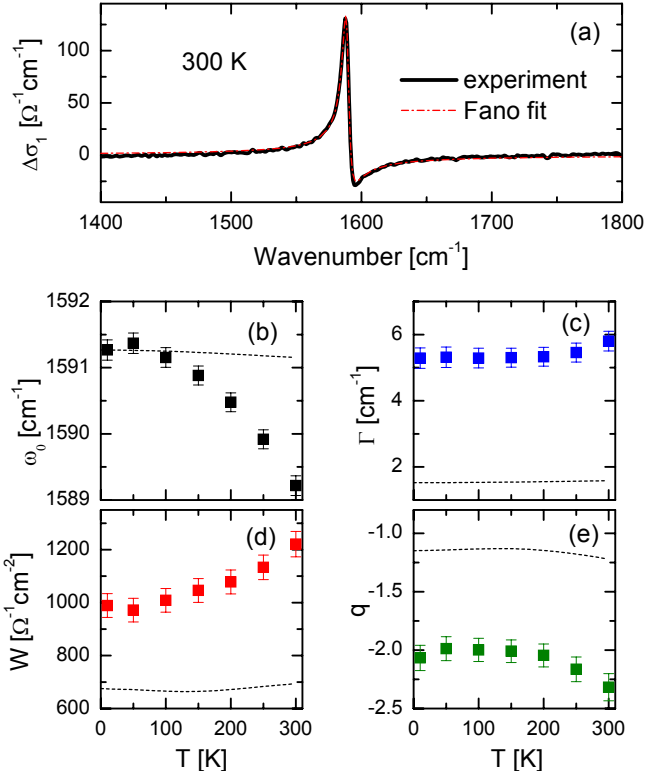}
\caption{(a) Real part of the optical conductivity
$\Delta \sigma_{1}(\omega)$ (thick black line)
in the vicinity of the $E_{u}$ phonon peak after a smooth background is subtracted.
Also shown is the fitting curve (red line) as obtained by
Eq. (\ref{eq:fano}).
Panels (b)-(e) show the phonon parameters $\omega_0$, $\Gamma$, $W$,
$q$ extracted by the fit as a function of temperature (symbols)
along with the theoretical calculations (dashed lines) not taking the phonon anharmonicity into account.}
\label{fig:fig2}
\end{figure}

Infrared reflectivity spectra were measured from 20 to 6400 cm$^{-1}$ at
different temperatures from 10 K to 300 K, using Fourier transform
spectroscopy and \emph{in-situ} gold coating as a reference. The spectra are
shown in Fig. \ref{fig:fig1}. The inset expands the region around 1600
cm$^{-1}$, where the phonon peak is located. Optical conductivity
$\sigma(\omega)$ was extracted by the Kramers-Kronig analysis, supported by
spectroscopic ellipsometry between 6400 and 36000 cm$^{-1}$ as described in
Ref. \onlinecite{kuzmenko2}. The spectral resolution was 5 cm$^{-1}$.
Fig. \ref{fig:fig2}a shows the real part of $\sigma(\omega)$ near the phonon
peak, where a smooth baseline due to direct electronic interband transitions
was subtracted. The most striking feature is a pronounced peak asymmetry,
signaling a Fano interference.\cite{fano,noi,tang,li} Accordingly, the
spectra at every temperature are fitted very well with a standard Fano
formula:\cite{fano,noi}
\begin{equation}
\Delta \sigma'(\omega) =
\frac{2W}{\pi \Gamma} \frac{q^2+2qz-1}{q^2(1+z^2)}
\label{eq:fano},
\end{equation}
where $z= 2(\omega-\omega_0)/\Gamma$, and where $\omega_0$ and $\Gamma$
characterize the frequency and the linewidth of the phonon anomaly, while $W$
and $q$ represent the phonon spectral weight and the Fano asymmetry
parameter, respectively. The temperature dependence of such phonon properties
as extracted from the fit procedure is shown in Fig. \ref{fig:fig2}(b)-(e).
The phonon frequency decreases by about 2 cm$^{-1}$ as the temperature
increases from 10 K to 300 K. The spectral weight $W$ increases from 950 to
1200 $\Omega^{-1}$cm$^{-2}$, which corresponds to the variation of the
infrared effective charge $e^{*} = Ze$ from 0.28 to 0.31 $e$, in a good
agreement with the previous experiments.\cite{nemanich,dresselhaus,humlicek}
The measured linewidth is close to 5 cm$^{-1}$, which means that it is probably
determined by the spectral resolution used and exceeds the actual value.

The observation of a finite infrared phonon intensity in pristine undoped
graphite is quite intriguing given that it is a monoatomic compound. No
infrared intensity would be expected for instance if all the carbon atoms were equivalent.
Moreover the evaluation of the static dipole in realistic graphite through a tight-binding
calculation finds, like in bilayer graphene, a charge-disproportion between
the inequivalent sites (and hence a static electric dipole) three order of
magnitude smaller than what observed.\cite{noi} A suitable framework to
account for the finite infrared activity is thus the charged-phonon theory,
which describes the borrowing of electronic effective charge by the optical
mode from the electronic transitions to which it becomes coupled.\cite{rice}
Such a theory has been successfully applied to bilayer graphene to explain the
strong dependence of the phonon intensity and Fano asymmetry as functions of
gating.\cite{capp}

A key role in this context is played by the mixed response function
$\chi(\omega)$,\cite{capp} which couples the
current operator with theelectron-phonon
scattering operator related with the $E_u$ lattice mode. The spectral weight is
given by the relation $W =\pi [\mbox{Re}\chi(\omega_0)]^2/\omega_0$, while
the asymmetry parameter is given by
$q=-\mbox{Re}\chi(\omega_0)/\mbox{Im}\chi(\omega_0)$.

\begin{figure}[t]
\includegraphics[scale=0.35,clip=]{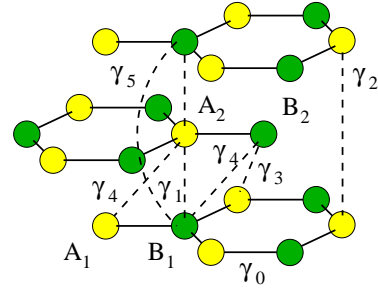}
\caption{(color online) (a) Atomic structure of graphite
showing the relevant hopping terms $\gamma_i$.
Atoms B1 and A2, connected by vertical $\gamma_1$,
denoted by darker colors, contain also a local
crystal field potential.}
\label{f-structure}
\end{figure}

The analysis in Ref. \onlinecite{capp} was performed within a simple
tight-binding (TB) model, where only the nearest-neighbor in-plane and
inter-plane hopping terms, $\gamma_0$ and $\gamma_1$, were considered
(see Fig. \ref{f-structure}).
Such
model can be generalized in a straightforward way for the bulk graphite by
introducing the interplane dispersion, so that $\chi(\omega) \rightarrow
\chi^{\rm 3D}(\omega) =\int_{-\pi}^\pi (dk_z/2\pi) \chi(k_z,\omega)$, where
$\chi(k_z,\omega)$ is the mixed response function for bilayer graphene with
the replacement $\gamma_1 \rightarrow \gamma_1(k_z)=2\gamma_1\cos(k_z /2)$.
Note however that the description of the electronic structure of graphite in
terms of this simple model would predict no infrared activity since the mixed
response function would be zero. As we will show, the inclusion of high-order
tight-binding terms, especially the ones that break the particle-hole
symmetry, is fundamental to understand the origin of the phonon activity.

In order to perform a quantitative analysis we consider the full
tight-binding model defined in the  $4 \times 4$ basis of the atomic orbitals
$ \Psi^{\dagger}_{\mathbf k} = (a^{\dagger}_{1 \mathbf k}, b^{\dagger}_{1
\mathbf k},  a^{\dagger}_{2 \mathbf k}, b^{\dagger}_{ 2\mathbf k}) $ where $
a^{\dagger}_{i \mathbf k}$ and $b^{\dagger}_{i \mathbf k} $ represent the
creation operators relative to the  A and B sublattices in the layer
$i$. 
The atomic basis and the different terms of the TB structure
are shown in Fig. \ref{f-structure}.
The
low energy bands of graphite around the K-H corner of the Brillouin zone are
described by the Hamiltonian
\begin{equation}
\hat{H}_{\bf k}
=
\left(
  \begin{array}{cccc}
    \tilde{\gamma}_{2}   & v\pi^{-}                     & \tilde{v}_{4}\pi^{-}         & \tilde{v}_{3} \pi^{+} \\
    v \pi^{+}       &    \tilde{\gamma}_{5} + \Delta &   \tilde{\gamma}_{1}        & \tilde{v}_{4} \pi^{-} \\
    \tilde{v}_{4}\pi^{+} & \tilde{\gamma}_{1}           & \tilde{\gamma}_{5}  + \Delta & v\pi^{-}              \\
    \tilde{v}_{3}\pi^{-} & \tilde{v}_{4}\pi^{+}         & v\pi^{+}                     & \tilde{\gamma}_{2}
  \end{array}
\right),
\label{grafite}
\end{equation}
where $\pi^{\pm} = \hbar (k_x\pm ik_y)$, $\tilde{\gamma}_{1,3,4} =
2\gamma_{1,3,4}\cos(k_{z}d/2)$, $\tilde{\gamma}_{2,5} = 2
\gamma_2+2\gamma_{2,5}\cos(k_{z}d)$, $\tilde{v}_{3,4} =
v\tilde{\gamma}_{3,4}/\gamma_{0}$. The paramagnetic current along the
$y$-axis is thus evaluated as $j_y = \sum_{\mathbf k, \sigma}
\Psi^{\dagger}_{\mathbf{k}} \hat{j}_{\mathbf{k}y} \Psi_{\mathbf{k}} $ where
$\hat{j}_{\mathbf{k}y} = -(e/\hbar) \partial \hat{H}_{\bf k}/\partial k_y $.
The coupled electron-phonon interaction operator can be written as $ V=
g\sum_{\mathbf k, \sigma} \Psi^{\dagger}_{\mathbf{k}} \hat{V}_{{\bf k}}
\Psi_{\mathbf{k}} \phi_y$ where $g$ is the electron-phonon coupling constant
as defined in Ref. \onlinecite{capp}, $\phi_y$ represents the dimensionless
in-plane lattice displacement along the $y$ direction. The precise form of
$\hat{V}$ depends on the TB model considered. In a simple model, where only
$\gamma_0$ and $\gamma_1$ are present, $\hat{V}_{{\bf k}}$ in graphite has
the same expression as derived in Ref. \onlinecite{ando} for bilayer
graphene. In a full TB model, however, $\hat{V}_{{\bf k}}$ requires the
knowledge of the derivative of $\gamma_4$ with respect to the lattice
distortion $u$, in addition to the well known term $d\gamma_0/du$. The matrix
$\hat{V}_{{\bf k}}$ thus reads
\begin{eqnarray}
\hat{V}_{{\bf k}} = \left(
\begin{array}{cccc}
 0  & 1 & - \alpha_4/\alpha_0 \Gamma  & 0 \\
 1  &  0 &  0 &  \alpha_4/\alpha_0 \Gamma \\
 -\alpha_4/\alpha_0 \Gamma  & 0 & 0 & -1 \\
 0  &  \alpha_4/\alpha_0 \Gamma   & -1 & 0
\end{array}
\right),
 \label{asymm_ph}
\end{eqnarray}
where $\Gamma = \cos(k_z d/2)$.
The deformation potentials $\alpha_0 = d\gamma_0/du= 4.4$ eV/\AA\,
 $\alpha_4 = d\gamma_4/du= 0.3$ eV/\AA\,
are extracted from LDA calculations.\cite{piscanec,lazzeri,capp2}

Once the current $j_y$ and the electron-phonon scattering operator $V$ are
defined, the phonon spectral properties can be evaluated in the framework of
the charged-phonon theory following the derivation for bilayer
graphene.\cite{capp} We evaluate the mixed response function in the
bare-bubble approximation, where we obtain the analytical expression:
\begin{eqnarray}
\chi(k_z, \omega)
=
\beta \sum_{{\bf k},n,m}
C^{nm}_{y,{\bf k}}
\frac{f(\epsilon_{n,\mathbf{k}}) - f(\epsilon_{m,\mathbf{k}})}
{\epsilon_{n,\mathbf{k}} - \epsilon_{m,\mathbf{k}} + \hbar \omega + i\eta},
\label{chi_ja_grafite}
\end{eqnarray}
where $n,m$ are band indices, $C^{nm}_{y,{\bf k}} =
(\hat{j}_{\mathbf{k}y})^{nm} (\hat{V}_{{\bf k}})^{mn} $ and $f(\epsilon)$ is
the Fermi function. We introduced a phenomenological damping term $\eta$ to
account for disorder effects. Finally, $\beta = -i g N_v N_s $, where $N_v =
2$ and $N_s = 2$ stand for the valley and spin degeneracy, respectively.

In contrast to the case of the simple $\gamma_0$-$\gamma_1$ model, the
inclusion of the higher-order TB terms makes $\chi(\omega)$ dependent on the
ultraviolet cutoff related to the finite bandwidth. To this aim we introduce
in the linearized model a momentum cutoff $k\le \bar{k}$, where $\bar{k}$ is
determined by requiring the conservation of the Brillouin zone, namely
$S_{\rm cell}=\pi N_K \bar{k}^{2}$. We get thus $\bar{k} = 1.095$ \AA$^{-1}$,
which is comparable with the distance $\Gamma$-K $\approx 0.98$ \AA$^{-1} $
of the hexagonal Brillouin zone.

To elucidate the role of the various TB  parameters on the infrared phonon
peaks, we consider for the moment the full set of TB parameters from
Ref. \onlinecite{alexey-kerr}. The phonon spectral properties obtained for this
model retaining all TB parameters, using $\eta=5$ meV and $T$=0 K, are
reported in Table \ref{tab:tab2}. As we can see, the inclusion of the
higher-order terms is crucial to give rise to a finite infrared activity.

\begin{table}[t]
\begin{tabular}{clcc}\hline\hline
TB model & p-h symmetry & $W$ & $q$ \\
\hline
full     & broken    & 666 & -1.15 \\
partial: $\gamma_2$        & broken    & 16  &  -0.53 \\
partial: $\gamma_3$        & preserved &  -  &  -      \\
partial: $\gamma_4$        & broken    & 471 & -0.94 \\
partial: $\gamma_5$        & broken    & 50  & -0.47   \\
partial: $\Delta$          & broken    & 332 & -0.74 \\
\hline\hline
\end{tabular}
\caption{The spectral weight $W$ and the asymmetry parameter $q$ at $T=0$ for
the full TB model and for the partial TB models described in the text. $W$ is
expressed in units of $\Omega^{-1}$ cm$^{-2}$.}
\label{tab:tab2}
\end{table}

In order to investigate the role of each TB parameter on the infrared
activity, we consider a set of toy TB model, based on our present parameter
set where, keeping the terms $\gamma_0$ and $\gamma_1$ fixed, we ``switch
on'', one by one, the remaining TB parameters $\gamma_2$, $\gamma_3$,
$\gamma_4$, $\gamma_5$ and the crystal field $\Delta$. The chemical potential
is evaluated self-consistently in each model to preserve the charge
neutrality. The results are summarized in Table \ref{tab:tab2}, where one can
clearly see how the onset of infrared activity is directly related to the
breaking of the particle-hole symmetry. Within this scenario, the detailed
optical properties result from a complex balance between all TB parameters.

%
\begin{figure}[b]
\includegraphics[scale=0.5,clip=]{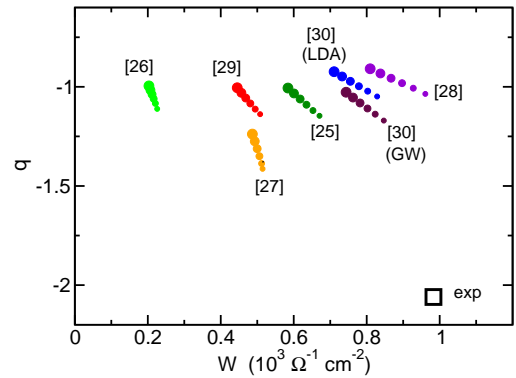}
\caption{
Phase diagram of the phonon strength $W$
vs. the Fano parameter $q$. The empty square represents
our experimental values,
filled circles represent data obtained at $T$ 10 K by using different TB models
available in literature.\cite{alexey-kerr,dresselhaus2,rabi,charlier,pp,attaccalite}
The size of symbols for the theoretical data is proportional to the
damping term $\eta=5,\ldots,30$ meV.}
\label{fig:fig3}
\end{figure}
To estimate the sensitivity of the predictions of the charged-phonon model on
the particular choice of the TB model, we calculated the phonon strength $W$
and the Fano factor $q$ obtained from different TB parameter sets proposed in
literature for
graphite.\cite{alexey-kerr,dresselhaus2,rabi,charlier,pp,attaccalite}
The results are
summarized in Fig. \ref{fig:fig3}, where we also show the experimental values
of $W$ and $q$ at 10 K. The large spread of the theoretical points in the
$W$-$q$ diagram reveals that relatively small changes in the underlying
electronic structure affect significantly the phonon properties. Given the
high sensitivity of the results of the charged-phonon model on the electronic
details and some uncertainty related to the choice of the momentum cutoff
$\bar{k}$, the comparison with the experimental finding seems to be
reasonably good. Importantly, the model predicts correctly the large value of
the experimental spectral intensity $W$ and therefore the effective infrared
charge. This is a significant improvement as compared to a previous calculation
based on a generalized bond-charge model,\cite{mahan} which predicts the
effective charge of this mode to be an order of magnitude smaller. The
charge-phonon theory also accounts for (and even overestimates) the strong
Fano lineshape asymmetry. It should be noted, however, that the precise value
of the Fano parameter $q$ is also quite sensitive to the damping magnitude,
which is sample dependent and should be regarded as an adjustable parameter
in the present model.


It is interesting to compare the temperature dependence of the phonon parameters shown in Fig.
\ref{fig:fig2} with the theoretical calculations (dashed lines). The frequency shift and the linewidth of the
phonon peak here are evaluated via the phonon self energy $ \Pi(\omega): \Delta
\omega_0 = \hbar^{-1} \mbox{Re}[\Pi(\omega_0)], \Gamma = -\hbar^{-1}
\mbox{Im}[\Pi(\omega_0)]$, while the spectral weight and the asymmetry parameter are obtained using the given above formulas.
We can notice that for all parameters the theory strongly underestimates the temperature variation as compared to the experiment.
We believe that the strong temperature dependence of the experimental features is a signature of
anharmonic effects induced by the phonon-phonon scattering. A careful
study of such anharmonic effects within the density functional theory context
was provided in Ref. \onlinecite{bonini}, where these affects were found to be mostly responsible for the thermal variation of the phonon frequency and the linewidth. It is likely that the phonon anharmonicity is also responsible for a sizeable temperature dependence of the spectral weight and the Fano asymmetry. The theoretical treatment of anharmonicity within the charged-phonon theory is an interesting problem which is beyond the scope of the present work.

In conclusion, we have investigated the origin of the infrared
phonon activity of the $E_u$ mode at $\omega \approx 1590$ cm$^{-1}$ in
pristine graphite. We experimentally found that the effective infrared charge is close to 0.3 electrons,
which is an exceptionally large value for an elemental compound.
A significant Fano asymmetry is also present ($q \approx$ -2). We have shown
that both these observations can be explained within the framework of the
charged-phonon theory, where a crucial role is played by the broken
particle-hole symmetry associated with the higher-order tight-binding terms.
The approach used in this work can be extended to other related materials, such as carbon
nanotubes and intercalated graphite.

We thank Lara Benfatto for many useful discussions. E.C. acknowledges the
Marie Curie grant PIEF-GA-2009-251904. The work of A.B.K. was supported by the grant No.200020-130093 of the Swiss National Science Foundation (SNSF).

\end{document}